\documentstyle[preprint,eqsecnum,aps]{revtex}

\begin{document}
\draft
\preprint{}
\title{ Delocalization of electrons  in  a  Random Magnetic Field}
\author{D.N. Sheng and  Z. Y. Weng}
\address{Texas Center for Superconductivity and Department of Physics\\
University of Houston, Houston, TX 77204-5506 }
\maketitle
\begin{abstract}
Delocalization problem for  a two-dimensional non-interacting
electron system is studied under  a random magnetic field.
With  the presence of  a random magnetic field, the  Hall conductance  carried
by each eigenstate can
become nonzero and quantized in units of $e^2/h$.  Extended states are
characterized by nonzero Hall conductance, and by studying finite-size scaling
of the density of extended states, an insulator-metal phase transition is
revealed. The metallic phase is found at the center of energy band which is
separated from the localized states at the band tails by critical energies $\pm
E_c$. Both localization exponent and the critical  energy $E_c$ are shown to
be dependent on the  strength of random magnetic field.
\end{abstract}

\vspace{0.8in}

\newpage
The Anderson localization theory$^{1,2}$ predicts that all states in a
two-dimensional (2D)  electron system are localized in the absence
of a magnetic field.
The quantum Hall effect (QHE) system is  a first  example of 2D systems which
show
the existence of truly extended states.$^{3,4}$ In this  latter case,
 the presence of a  magnetic
 field breaks the time-reversal symmetry and destroys
constructive interference of  the backward scattering$^{2}$ so that  it is
  possible for electrons to propagate forwardly.

Recently, an intensive attention has been attracted to the delocalization
problem in a 2D random-magnetic-field system. This problem is closely
related to the half-filled QHE system$^{5,6}$ as well as the gauge-field
description$^{7,8}$ of the high-T$_c$ superconductivity problem. However,
despite a lot of
numerical and  theoretical  efforts, the issue of delocalization still remains
 controversial.  Theoretically, Zhang and Arovas$^{9}$ have recently argued
that the field-theory description, which  corresponds to a non-linear
sigma-model of the  unitary class without  a topological term due to zero
 average of magnetic field, should have  a  term representing a
 long-range logarithmic interaction  of the topological density ( due to the
local magnetic field). This singular term may lead to a phase transition
from localized state to extended one. But it is contradictory to the
 conclusion that all the states are localized obtained by
Aronov, Mirlin and Wolfle$^{10}$ in a similar approach. Earlier numerical
works$^{11-13}$ also have given conflicting results. Recently, with a larger
sample size, Liu {\it et al.}$^{14}$ have found a scaling behavior of the
localization length near the energy band tail, which can be extrapolated to
give a insulator-metal transition energy $E_c$. Nevertheless,
 a metallic phase has not been directly confirmed since no scaling behavior
 has been found there.  In the  possible metallic region, an even larger
sample size may be needed in order to distinguish whether  the states are
 really extended or very weakly localized$^{11}$  with
the localization length much longer than the sample size.

Thus it would be desirable to study this
delocalization problem from an alternative numerical method which directly
probes topological properties of system with less finite-size effect.
Thouless and co-workers$^{15}$ and others $^{16,17}$
have found that delocalization property of a wavefunction in the presence of
magnetic field can be well characterized by its associated quantized Hall
conductance.  Nodes of an eigenstate wavefunction
with nonzero Hall conductance  can move freely and cover
the whole real space when one continuously  change the boundary
condition.$^{17}$ Such a covering of real space by the nodes
has been related to  a topological invariant
integer (known as  First Chern number ), which is identical to the quantized
number of the Hall conductance (in unit of $e^2/h$). Thus a nonzero Hall
conductance describes  the extensiveness of a wavefunction.
In contrast, a zero-Hall conductance state will always be localized in 2D
with the presence of weak-impurities (Anderson localization).
In the QHE system, Huo and Bhatt$^{18}$ have calculated
the boundary-phase-averaged Hall
conductance for each eigenstate of a noninteracting electron system in the
presence of strong magnetic field, and extrapolated the density of
extended states (with nonzero Hall conductance) to the thermodynamic
 limit ( sample size varying from 8 to 128). They have found that all extended
states collapse to a single energy $E_c$ at the center of the Landau band,
with a localization length  $ \xi \propto 1/|E-E_c|^{\nu}$ and localization
exponent $\nu=2.4 $  in agreement with previous known  results.

In the present random magnetic field case,  the total Hall conductance
on average has to be zero. But one  still finds nonzero quantized
Hall-conductances for eigenstates at each random flux configuration.
Due to the general relation between a nonzero Hall
conductance and delocalization of the corresponding wavefunction,$^{15}$
one can use this topological quantity to characterize delocalized states.
Similar point of view also lies in the heart of the field-theory approach
 of Zhang and Arovas.$^9$
In this Letter, we shall use this topological property in our numerical
approach. By studying the sample-size dependence of the density of extended
states which are states with nonzero Hall conductance,
 an insulator-metal phase transition will be revealed.
The extended states are found  near the center of energy band,
 and the states at the band tail are all localized
with both localization exponent and transition energy $\pm E_c$ depending on
the random magnetic field strength.

We consider a tight-binding lattice model   of   noninteracting electrons
under a random magnetic  field. The Hamiltonian is defined as follows:
$$H=-\sum_{<ij> } e^{i a_{ij}}c_i^+c_j +\sum _i w_i c^+_i c_i
\eqno (1)
$$
Here $c_i^+$ is a  fermionic creation operator, with $<ij>$
referring to two  nearest neighboring sites.  A magnetic flux per plaquette is
given as  $\phi=\sum _ {\Box} a_{ij}$, where the  summation runs  over four
links around a plaquette. We study the case in which $\phi$ for
each plaqutte is randomly distributed between $-h_0 \pi $ and $h_0 \pi $, and
$w_i$ is also a random potential with strength $|w_i|\leq w$. For simplicity,
we assume no correlations  among different plaquettes for $\phi$ and different
sites for $w_i$ (white noise limit).    The total flux for each random
configuration is always chosen to be zero.
The finite system  is diagonalized under the generalized boundary condition
 $|\Psi (i+L_j)>=e^{i \theta _j}|\Psi (i)> $
 (j=1,2 represent x and y direction respectively) with lattice width $L_1=
L_2=L$, and a total number of lattice sites (sample size) is
  ${\cal N}=L \times L$ (the lattice constant is chosen to be the unit).

The Hall conductance     can be  calculated by using the Kubo
formula.  One may relate a Hall conductance to each eigenstate $|m>$:
$$ \sigma _{xy}^{(m)}=\frac {i e^2 h}{2 \pi } \sum_{n\neq m} \frac
{<m|p_x|n><n|p_y|m>- <m|p_y|n><n|p_x|m>}
 {(\varepsilon_m-\varepsilon_n)^2}    \eqno(2) $$
where ${\bf p} $ is the velocity operator defined as
$p_{\tau}=i \sum_{i}(c_{i+\tau}^+c_{i}e^{i a_{i+\tau,i}}
-c_{i}^+c_{i+\tau}e^{-i a_{i+\tau,i}})  $ with $\tau={\hat x} $ or ${\hat y}$.
The total Hall conductance for the system is given by $\sigma _{H} =\sum _{
\varepsilon_m < \varepsilon _F } \sigma _{xy} ^{(m)}$ at zero temperature,
with $\varepsilon _F  $ as the Fermi energy.
$\sigma_H$ will always be zero on average in the case of a random magnetic
field. However,  $\sigma_{xy}^{(m)}$ can be  nonzero for each random-flux
configuration because of the  breaking of time-reversal symmetry.
As pointed out before, a state with nonzero (quantized) Hall conductance
represents an extended state in the thermodynamic limit, whereas a
zero-Hall conductance state should always be localized in 2D.

Direct calculation of matrix elements in formula (2) is  time consuming.
We can make a unitary transformation $|\Phi >=e^{-i \theta _1x/L}
e^{-i \theta _2 y/L} |\Psi> $ such that the new state $|\Phi> $  satisfies a
periodical boundary  condition.  The Hamiltonian (1) is  transformed
in terms of $ a_{ii+ {\tau }}  \rightarrow   a_{ii+{\tau}}+
\theta _{\tau}/L $ with $\tau=1({\hat x}) $ or 2 (${\hat y})$.
Then the boundary-phase averaged conductivity can be  related to partial
derivatives of the wave function in the
following form:$^{15}$
$$ \sigma _{xy}^{(m)}=\frac {ie^2} {4 \pi h }  \oint d \theta _j \sum
_{i}
\left( \Phi _m^* (\theta,i)\frac {\partial \Phi _m (\theta,i)}
{\partial \theta _j}
 - \frac {\partial \Phi ^* _m (\theta,i)}  {\partial \theta _j} \Phi _m
(\theta,i)\right)
  \eqno (3)$$
where the closed-path of the integral is along the boundary of a unit cell
$0\leq \theta_1, \theta_2 \leq 2\pi$.
$\sigma_{xy}^{(m)}$ in (3) can be shown$^{15,16}$ to be quantized in unit of
$e^2 /h$. Here $\Phi$ is required to be an analytic  wave function in 2D
$\theta$-space.
Starting from the wavefunction $\Phi (0,0) $ defined at
one  corner of the boundary in $\theta$-space,  the
phase of the wavefunction $\Phi$ can  be uniquely determined$^{15}$
by a process of parallel translation, first along $\theta _1 $ axis and
then along  $\theta_2 $ axis as:
$$  \sum _i \Phi ^{*} (\theta _1 , 0) \frac {\partial  \Phi (\theta
_1,0) } {\partial \theta _1 } =0,     \eqno(4a)            $$
$$  \sum _i \Phi ^{*} (\theta _1 , \theta _2) \frac {\partial  \Phi (\theta
_1, \theta _2)} {\partial \theta _2 } =0.     \eqno(4b) $$

Numerically we have diagonalized the Hamiltonian  with  boundary angle
varying in whole $2 \pi \times 2\pi $ phase space for
each given  random flux and potential configuration.  At each step,
$ \theta _j $ may only change by a very small value  such that $ \partial
\Phi /\partial \theta$ can be well approximated
by       $ \Delta \Phi /\Delta \theta $ (Usually $ \Delta \theta
< 2 \pi /100 $ which is adjustable in our numerical calculation to give a
reliable result). By constructing a wavefunction satisfying conditions (4),
the Hall conductance averaged over boundary angle is determined in terms of (3)
for each eigenstate.
An eigenstate with nonzero Hall conductance is defined as an extended state,
and the corresponding  density of states  $\rho _{ext} (\varepsilon, {\cal N} )
$ is obtained
as a function of energy $\varepsilon$ and sample size ${\cal N}$ which is
averaged over random flux-potential configurations ($200-2000$ random
configurations depending on sample size).

The total density of
states  $\rho (\varepsilon , {\cal N})$  and the extended one
$\rho _{ext}(\varepsilon , {\cal N})$ are obtained  as a function
of energy $\varepsilon $ and  lattice size ${\cal N}$ ( ${\cal
N}$=16,36,64,100 and 144).
The total density of states does not change  much with lattice size,
but the extended part of the density of states shows distinctive behaviors at
different energy regions separated by  critical energies $\pm E_c$.
The ratio $\rho _{ext }(\varepsilon,{\cal N})/ \rho (
\varepsilon, {\cal N})$ is presented in Fig. 1 around
$-E_c$ (with
 random magnetic field  and impurity strengths chosen as $h_0=0.6\pi $ and
$w=1.0$, respectively).
 All the curves   in Fig. 1 cross at a fixed-point $ \varepsilon= -E_c $, which
is independent of the lattice size. At energy $\varepsilon < -E_c $,  the
extend-state density is continuously suppressed, and can be
extrapolated down to zero as lattice size becomes infinity (see below). On
the other hand, in the regime  $-E_c < \varepsilon < E_c $,  $\rho _{ext}/\rho$
monotonically increases
with lattice size and eventually saturates.  Therefore,  Fig. 1 clearly
shows a metal-insulator transition at  critical energies $\pm E_c$
(the curves in Fig. 1  are  symmetric about $\varepsilon =0$).

Let us consider  in detail the localization at the band tail
$\varepsilon < -E_c$. One may define two quantities
characterizing the localization effect:
a  ratio $R_0$ of the number of extended states divided by the total number of
states  at  energy $\varepsilon < -E_c $ region, and a mean width
$\Delta E$ of the extended states  in such a regime, both of which presumably
 will approach to zero in
the thermodynamic limit. Here
$R_0= \int _{-\infty}^{-E_c}
\rho _{ext}(\varepsilon) d\varepsilon/
 \int _{-\infty}^{-E_c}
\rho (\varepsilon) d\varepsilon $  and
$(\Delta E)^2 =  \int _{-\infty}^{-E_c}
[\varepsilon -(-E_c )]^2 \rho _{ext}(\varepsilon) d\varepsilon
/ \int _{-\infty}^{-E_c} \rho _{ext}(\varepsilon) d\varepsilon$.
$R_0$ and $\Delta E$ versus the sample size are shown in Fig. 2 in a log-log
plot at $h_0=0.6$ and $w=1.0$.
The data follow two parallel straight lines nicely, suggesting the following
power-law behavior:
$R_0 \sim {\cal N}^{-x} $ and $\Delta E \sim {\cal N}^{-x} $,
with $x=0.2\pm 0.02 $. Such a scaling law ensures the absence of
the extended states outside the energy range ($-E_c$, $E_c$) in the
thermodynamic
limit.   In the localized region,  localization length is a characteristic
length scale ( scaling parameter$^{19}$),
and for a finite-size sample with a width  $L$ the
states with a localization length $\xi  >  L $ should appear as
extended ones. If  the localization length goes as
$1/|\varepsilon-(-E_c)|^{\nu}$ when $\varepsilon \rightarrow -E_c$,
 one expects
$1/(\Delta E)^{\nu} \sim L$, or $\Delta E \sim {\cal N}^{-1/2\nu}$ (Ref. 18).
One also has $R_0\propto \Delta E \rho(-E_c, {\cal N})\propto \Delta E$.
So the finite-size scalings of $R_0$ and $\Delta E$ in Fig. 2 imply a
power-law  behavior of the localization length $\xi $: $\xi
 \propto 1/|\varepsilon -
(-E_c) |^{\nu}$ with $\nu=1/2x=2.5 \pm 0.3 $.

At $-E_c < \varepsilon < E_c $, a monotonic increase of $\rho _{ext}/ \rho $
with  sample size is manifestly  metallic behavior. It is consistent with the
behavior of $d (\xi _L/L )/d L >0 $
( $ \xi _L $ is the so-called decay length and $\xi _L/L$  describes the
extensiveness of the system) found in the metallic region of 3D system
and 2D system with spin-orbit interaction (symplectic class).$^{19,20  }$
In the present approach, the quantity
 $\rho _{ext}$ directly characterizes extended  states and can be
extrapolated to  a finite value at large sample size
limit.  One may also define a ratio
$N_{ext}/N_0=\int^{E_c}_{-E_c} \rho_{ext}d\varepsilon/\int^{E_c}_{-E_c} \rho
d\varepsilon$, namely  the total number of extended states divided by the
total number of states  within ($-E_c$, $E_c$). $N_{ext}/N_0$ is found
to saturate to a finite value $R_c\sim
0.68$ in the following  manner: $N_{ext}/N_0-R_c\propto -N^{-y}$ ($y\sim 0.3$),
which is shown in Fig. 3 by a log-log plot ($h_0=0.6$, $w=1.0$). A finite $R_c$
in the thermodynamic limit is a direct evidence for delocalization.

Very similar behaviors have also been obtained at
other random-flux strengths: $h_0= 0.4$ and $0.5$ (with $w=1.0$).
Correspondingly,  $E_c=\pm 3.7$ and $\pm 3.5$,  while $\nu = 1.25 \pm 0.3$ and
 $1.75\pm 0.3$, respectively. The results suggest a non-universal localization
exponent $\nu$,  which increases with $h_0$ and is consistent with a larger
$\nu$ ($\sim 4.5$) obtained at $h_0>0.7$ in Ref. 14.
The reduction of the metallic region ($-E_c, E_c$)  indicates
that the extended states are less favorable   at larger $h_0$.
  With the increase of
$h_0$ we find that $\rho_{ext}$ becomes less sensitive to the sample size.
When $h_0>0.7$, $\partial\rho_{ext}/\partial N$ is relatively small
around the center of the energy band  and a
larger lattice size is needed in order to get  conclusive results about
delocalization.
Since  we study the density of states for
the extended states  characterized by the topological properties of
wavefunctions (Hall conductances), finite-size effect is expected to be less
important here in comparison with other approaches.  The existence of
the fixed-points $\pm E_c$, which are independent of  lattice size, as
well as the finite-size scalings on two sides of $\pm E_c$ indeed support this
expectation.

In conclusion,  we have unambiguously demonstrated
the existence of a delocalization  region for  a
non-interacting 2D electron system  under a random magnetic field. Critical
energy $E_c$
of metal-insulator transition has been  determined. Two branches of
finite-size scaling are found in both metallic and localized regions, and the
results are extrapolated to the thermodynamic limit.  The localization
length at  the band tail
($\varepsilon < -E_c$, $\varepsilon > E_c$) behaves like $ \xi \sim
1/|\varepsilon \pm  E_c|^{\nu}$,
with both $E_c$ and $\nu$ varying with the strength of  random magnetic field.

{\bf Acknowledgments} -The
present work is supported  by the Texas Center for Superconductivity at the
University of Houston and  grants from Robert A. Welch Foundation and Texas
Advanced Research Program.

\newpage

\noindent $^1$ E. Abrahams, P. W. Anderson, D. C. Liciardello, and V.
Ramakrishnan, Phys. Rev. Lett. {\bf 42}, 673 (1979).\\
$^2$ For a review, see, P. A. Lee and T. V. Ramakrishnan, Rev. Mod. Phys.
${\bf 57}$, 287 (1985).\\
$^3$ R. B. Laughlin, Phys Rev. B. ${\bf 23}$, 5632 (1981); B. I. Halperrin,
Phys. Rev. B. ${\bf 25}$, 2185 (1982).\\
$^4$ H. Levine, S. B. Libby and A. M. M. Pruisken, Phys. Rev. Lett. ${\bf 51}$,
1915 (1983); A. M. M. Pruisken, in {\bf The   Quantum  Hall  Effect}, edited
by R. E. Prange and S. M. Girvin ( Springer-Verlag, Berlin, 1990).\\
$^{5}$V. Kalmeyer and S. C. Zhang, Phys. Rev. B. ${\bf 46}$, 9889 (1992).\\
$^{6}$B. I. Halperin, P. A. Lee and N. Read, Phys Rev. B. ${\bf 47}$, 7312
(1993).\\
$^{7}$ L. B. Ioffe and A. I. Larkin, Phys. Rev. B ${\bf 39}$, 8988 (1989);
N. Nagaosa and P. A. Lee, Phys Rev. Lett. ${\bf 64}$, 2450 (1990); Phys. Rev.
B. ${\bf 45}$, 966 (1992).\\
$^{8}$ Z. Y. Weng, D. N. Sheng and C. S. Ting, TCSUH preprint
no. 95:011 (1995).\\
$^{9}$ S. C. Zhang and D. Arovas, Phys. Rev. Lett. {\bf 72},
1886, (1992).\\
$^{10}$ A. G. Aronov, A. D. Mirlin and P. Wolfle, preprint (1994).\\
$^{11}$ T. Sugiyama and N. Nagaosa, Phys. Rev. Lett. ${\bf 70}$, 1980 (1993).\\
$^{12}$ Y. Avishai and Y. Hatsugai and M. Kohmoto, Phys. Rev. B. ${\bf 47}$,
9561 (1993). \\
$^{13}$  V. Kalmeyer, D. Wei, D. P. Arovas, and  S. C. Zhang,
Phys. Rev. B. ${\bf 48}$, 11095 (1993).\\
$^{14}$ D. Z. Liu, X. C. Xie, S. Das Sarma   and  S. C. Zhang,
 preprint (1994).\\
$^{15}$ D. J. Thouless, M. Kohmoto, M. P. Nightingale, and M. den Nijs, Phys.
Rev. Lett. ${\bf 49}$, 405 (1982); Q. Niu, D. J. Thouless, and Y. S. Wu, Phys.
Rev. B {\bf 31}, 3372 (1985),
 D. J. Thouless, J. Phys. C, {\bf 17}, L325 (1984).\\
$^{16}$ M. Kohmoto, Ann. Phys. (NY) ${\bf 160}$, 343 (1985).\\
$^{17}$ D. P. Arovas et. al., Phys. Rev. Lett. ${\bf 60}$, 619 (1988).\\
$^{18}$ Y. Huo and R. N. Bhatt, Phys. Rev. Lett. ${\bf 68}$, 1375 (1992).\\
$^{19}$ A. MacKinnon and B. Kramer,
Phys. Rev. Lett. ${\bf 47}$, 1546 (1981); Z. Phys. B ${\bf 53}$, 1  (1983).\\
$^{20}$  T. Ando, Phys. Rev. B {\bf 40}, 5325 (1989).

\newpage
\begin{center}
{\bf FIGURE CAPTION}
\end{center}

Fig. 1.  The ratio of the density of extended states over  the total
 density  of states   is  plotted
as a function of energy (around $-E_c=-3.3$)
 for different sample size ${\cal N}$. \\

Fig. 2. $R_0$, the  number of  extended states divided by the total number of
states at  ( $\varepsilon < -E_c $),  and  $\Delta E$,
the mean  width of the band of extended states
  in the same energy region,  vs. sample size ${\cal N}$ on a
log-log scale.   $E_0$
is a  mean width of the  band of  total states  at ( $\varepsilon < -E_c $).
  All the data are fit
into two parallel straight lines.                                \\

Fig. 3. $N_{ext}/N_0$, the number of extended states over
the  total  number of states within energy region (-$E_c$, $E_c$),
  vs. sample size ${\cal N}$.
The ratio is extrapolated to a finite number $R_c=0.68$ in the thermodynamic
limit,    and solid line is a best fit to  the data.                  \\

\end{document}